\documentclass[12pt]{article}
\usepackage[margin=1.25in]{geometry} 
\usepackage{amsmath,amsthm,amssymb}
\usepackage{multirow}
\usepackage{times}
\usepackage{listings}
\usepackage{graphicx}
\usepackage{enumerate}
\usepackage[colorlinks,
linkcolor=blue,
anchorcolor=blue,
citecolor=blue
]{hyperref}
\usepackage{dsfont}
\usepackage{float}
\usepackage[title]{appendix}
\usepackage{natbib}

\usepackage{setspace}
\makeatletter

\newcommand{\Rmnum}[1]{\expandafter\@slowromancap\romannumeral #1@}
\makeatother
\allowdisplaybreaks

\newtheorem{theorem}{Theorem}

\newtheorem{lemma}{Lemma}

\newtheorem{example}{Example}

\title{Ambiguous Persuasion: An Ex-Ante Formulation\footnote{This paper is a revised version of a chapter in my Northwestern University doctoral dissertation. I am deeply indebted to Peter Klibanoff and Marciano Siniscalchi for their guidance and support. Suggestions from the editor and three anonymous referees have greatly improved the paper. I also thank Modibo Camara, Andres Espitia, Sidartha Gordon, Yingni Guo, Jian Li, Ming Li, Alessandro Pavan, Harry Pei, Henrique Brasiliense de Castro Pires, Ludvig Sinander, and Udayan Vaidya for their comments.}}

\author{Xiaoyu Cheng\footnote{Department of Economics, Florida State University, Tallahassee, FL, USA. E-mail: \href{mailto:xcheng@fsu.edu}{xcheng@fsu.edu}.}}
\begin{document}
	\maketitle
	\begin{abstract}
		Consider a persuasion game where both the sender and receiver are ambiguity averse with maxmin expected utility (MEU) preferences and the sender can choose an ambiguous information structure. This paper analyzes the game in an ex-ante formulation: the sender first commits to an information structure, and then the receiver best responds by choosing an ex-ante message-contingent action plan. Under this formulation, I show it is never strictly beneficial for the sender to use an ambiguous information structure as opposed to a standard unambiguous one. This result is robust to (i) the players having heterogeneous beliefs over the states, and/or (ii) the receiver having non-MEU, uncertainty-averse preferences. However, it is \emph{not} robust to the sender having non-MEU preferences.

		
	\end{abstract}

	\newpage
	\section{Introduction}
	Strategic information provision, or more specifically persuasion, is valuable for an informed sender in influencing the action taken by a receiver \citep{bergemann2019information,kamenica2019bayesian}. The sender can influence the receiver's decision by providing information that shapes the uncertainty the receiver is facing. The literature has focused primarily on Bayesian persuasion in which the sender is only allowed to expose the receiver to uncertainty that can be fully described by a probability measure. For decisions under uncertainty, however, ambiguity aversion is recognized as an important phenomenon when facing uncertainty that is not describable by a probability measure, and it can influence behaviors (see a survey by \citet{gilboa_marinacci_2013}). This raises the question in the context of persuasion, first addressed by \citet*{beauchene2019ambiguous} (BLL henceforth): If the receiver is known to be ambiguity averse, can the sender benefit from strategically introducing ambiguity into their communication to better influence the receiver's decisions? 
	
	Indeed, BLL answer this question by studying a persuasion game under the assumption that both the sender and receiver are ambiguity averse with preferences represented by the Maxmin Expected Utility (MEU) model \citep{gilboa1989maxmin}.\footnote{In their paper, they also discuss possible extensions to more general ambiguity averse preferences.} In addition, the sender can choose to commit to an ambiguous information structure, an \textit{ambiguous experiment}, which specifies a closed and convex set of probability measures over states and messages shared by both players. The receiver observes the sender's choice of experiment and a realized message, then chooses an action according to their updated preferences. Among other results, they demonstrate that the sender can strictly benefit from choosing an ambiguous experiment and provide methods to characterize its extent.\footnote{See \cite{cheng2021concavification} for an alternative characterization using concavification.}
	
	This paper analyzes the persuasion game under the same assumptions as in BLL but adopts an \textbf{ex-ante formulation}: after observing the sender's choice of experiment, but \textit{before} any message is realized, the receiver chooses a message-contingent action plan (this is referred to as the \textit{ex-ante} stage for the receiver). The main result, Theorem \ref{ex-ante}, shows that under this ex-ante formulation, the sender can never strictly benefit from introducing ambiguity into their communication. In other words, a seemingly small difference in formulations of the game leads to a dramatic change in perspective on the value of ambiguous communication to the sender.
	
	These contrasting conclusions arise from the well-known non-equivalence between ex-ante and interim decisions made by an MEU decision maker when preferences are updated using any consequentialist updating rule, including the one assumed in BLL \citep*{hanany2007updating, hanany2009updating, siniscalchi2009two}. Formally, such a receiver is \textit{dynamically inconsistent}: their interim optimal actions may differ from those prescribed by their ex-ante optimal action plan. While BLL's persuasion game models the receiver as choosing their interim optimal action, the ex-ante formulation models the receiver as following their ex-ante optimal plan. To build some intuition for why this difference in modeling can lead to such divergent conclusions, consider the following example.
	
	\begin{example}\label{exp0}
		Suppose there are two equally likely states of the world: $\{\omega_{1}, \omega_{1}\}$. A representative voter (the receiver) is choosing between two policies: \textit{\{Status Quo, New\}}. The Status Quo is considered the safer action as its outcomes are known, while the New policy may be better or worse depending on the state. Formally, the voter's payoffs are given by the following payoff matrix: 
		\begin{center}
			\begin{tabular}{ccc}
				& $\omega_{1}$ & $\omega_{2}$  \\
				\hline
				Status Quo & 0 & 0 \\
				New & 1 & -0.5  \\
			\end{tabular}
		\end{center}
		A politician (the sender), who benefits from the Status Quo, strictly prefers the voter to choose it in all states. They can design and commit to an experiment (e.g., a policy experiment) to influence the voter's choice. 
	\end{example}

	Without any additional information, the receiver strictly prefers New to Status Quo as the states are equally likely. Under Bayesian persuasion, the best the sender can achieve is to induce the receiver to choose Status Quo three quarters of the time.\footnote{This is obtained by committing to a probabilistic information structure, a statistical experiment with two messages such that the receiver's posterior given these two messages are $0$ and $2/3$ in terms of the probability of $\omega_{2}$. The probabilities of sending these two messages are $1/4$ and $3/4$ respectively.}
	
	When both players have MEU preferences and the receiver chooses their interim optimal action, BLL show that the sender can induce the receiver to \textit{always} choose Status Quo using an ambiguous experiment. Specifically, the sender can commit to fully revealing the state but be ambiguous about which message corresponds to which state. Upon observing any message, if the receiver take both possibilities into account, their updated belief is fully ambiguous about the state.\footnote{This holds when the receiver applies full Bayesian updating as assumed in BLL.} As a result, under the MEU criterion, taking the Status Quo becomes the receiver's interim optimal given such an updated belief. 
	
	The receiver's interim optimal actions correspond to the message-contingent action plan that prescribes choosing Status Quo after every message. This yields the receiver an ex-ante payoff of zero. In contrast, the plan of choosing New after every message delivers a strictly positive ex-ante payoff under the same ambiguous experiment.\footnote{This follows from the fact that the receiver's ex-ante payoff from this plan is the same and strictly positive under any state-revealing statistical experiment.} Therefore, if the receiver seeks to maximize their ex-ante payoff, the plan of choosing Status Quo cannot be induced by the ambiguous experiment. Importantly, in this case, using the ambiguous experiment makes the sender's strictly worse off than under Bayesian persuasion. To summarize, the sender's optimal payoff in BLL's persuasion game relies on inducing message-contingent action plans that are not implementable under the ex-ante formulation. 
	\bigskip 

	Theorem \ref{thm:robustness} shows the conclusion of Theorem \ref{ex-ante} continues to hold when (i) the players having heterogeneous beliefs over the states, and/or (ii) the receiver is an Uncertainty Averse decision maker as defined in \cite{cerreia2011uncertainty}. However, one direction in which Theorem \ref{ex-ante} is not robust, is regarding the assumption the sender's preferences are MEU. In a follow-up paper, \citet{cheng2023} demonstrate that a non-MEU ambiguity averse sender may obtain strict benefits from ambiguous communication, even when all other assumptions of Theorem \ref{ex-ante} are preserved. While a full characterization of when and how the sender can benefit from ambiguous communication is provided in \citet{cheng2023}, the present paper includes a simple example (Example \ref{exp}) to illustrate the possibility of such benefits for a non-MEU sender.

	I close the introduction with a brief discussion on the importance of studying ambiguous persuasion also under the ex-ante formulation. Unlike in Bayesian persuasion, where the ex-ante and interim formulations are equivalent, analyzing persuasion games under non-Bayesian preferences, such as ambiguity aversion, requires careful consideration of both perspectives. This paper highlights the dramatic differences in the sender's optimal payoffs across the two formulations under ambiguity aversion. Thus, a comprehensive understanding of ambiguous persuasion requires examining both perspectives. 

	While both perspectives are equally important, the ex-ante formulation may be more appropriate in certain contexts. A more detailed discussion appears in Section \ref{discussion}, but here I briefly outline two scenarios. First, when the sender communicates through a contingent contract and the receiver is legally required to commit at the ex-ante stage, the ex-ante formulation might be more suitable. Second, the receiver may adopt a dynamically consistent updating rule, as in \citet{hanany2007updating, hanany2009updating}, which ensures that their interim choices align with their ex-ante plan.
    
    As a concrete illustration, \citet{SHISHKIN2023105610} conduct a lab experiment examining subject's response to a decision problem and an ambiguous experiment identical to those described in Example \ref{exp0}. They find that a majority of the subjects choose the ex-ante optimal action. This behavior suggests that, when such subjects serve as receivers, a sender may fail to realize the gains from ambiguous persuasion identified by BLL, and could even do worse than using a standard statistical experiment---emphasizing the behavioral plausibility and practical relevance of the ex-ante formulation. 

	The remainder of this paper is organized as follows. Section \ref{literature} reviews the related literature. Section \ref{dc} introduces the ambiguous persuasion game. Section \ref{nogain} presents the main results. Section \ref{discussion} provides additional discussions on the relevance of the ex-ante formulation. 

	\subsection{Related Literature}\label{literature}
	This paper contributes to the literature on Bayesian persuasion initiated by \cite{kamenica2011bayesian}, and more specifically, the line of research combining persuasion and ambiguity aversion \citep*{beauchene2019ambiguous, hedlund2020persuasion, NIKZAD2021144,cheng2022relative,kosterina2022persuasion}.	While this paper shows that an MEU sender cannot benefit from ambiguous persuasion in the ex-ante formulation, in a follow-up paper, \citet{cheng2023} characterize when and how a non-MEU sender may benefit.\footnote{\citet{cheng2023} extend the analysis to the setting in which both players have smooth ambiguity preferences \citep{klibanoff2005smooth}, of which MEU is a special case.} The ex-ante formulation adopted by the present paper, where the receiver chooses a message-contingent plan has also been used in other contexts, for example, the study of belief polarization \citep{baliga2013}, information order \citep{LI201618, LI2020105012,wang2023informativeness}, and incomplete information games \citep{hanany2020incomplete}. 
	
	The issue of dynamic consistency in ambiguous persuasion is also discussed in BLL and \citet{pahlke2022dynamic}. But they take different views than the present paper. They consider dynamic consistency through the lens of rectangularity \citep{epstein2003recursive}, which is a restriction imposed on the ambiguous experiment (effectively the set of distributions over states and messages), that, if satisfied, ensures the MEU receiver is dynamically consistent under full Bayesian updating. However, not all ambiguous experiments satisfy such a constraint, for example, those used in Examples \ref{exp0} and \ref{exp} are not rectangular. BLL show that if the sender is \textit{constrained} to only rectangular ambiguous experiments, then there is no gain for the sender. The nature of their result is fundamentally different from the main result of this paper, as the sender here is not constrained when choosing experiments. 
	
	\citet{pahlke2022dynamic} incorporates the idea of rectangularity by introducing correlations between the realization of states and which statistical experiment is used to generate the message. By doing so, the receiver's interim optimal actions will remain optimal according to the ex-ante belief with correlation, as it effectively enlarges the set of priors to its rectangular hull. As a result, in her paper, the sender's optimal payoff from ambiguous persuasion is the same as in BLL. In contrast, this paper studies the receiver's decision based on their original ex-ante belief without introducing such correlations. 
	
	There is a connection of this paper with papers on Bayesian persuasion when the players' priors are heterogeneous \citep*{alonso2016bayesian, laclau2017public, galperti2019persuasion}. Given an ambiguous experiment, because the sender and receiver's utility functions are different, the minimizer of their ex-ante payoffs may turn out to be different thus as if they have different priors. However, notice that this heterogeneity arises endogenously, while all the listed papers consider exogenously heterogeneous priors. Therefore, the results and findings of this paper cannot be obtained as implications of those papers. In addition, the main result of the paper is also shown to be robust to the players having heterogeneous beliefs over the states.
	
	\section{The Ex-Ante Formulation of the Persuasion Game}\label{dc}
	Consider a persuasion game between a sender and a receiver as in BLL, but in an ex-ante formulation. Let $\Omega$ be a finite set of \emph{states of the world}. For any finite set $X$, let $\Delta(X)$ denote the set of all probability distributions over $X$ endowed with the topology of weak convergence. The sender and receiver have a common prior $p \in \Delta(\Omega)$ with full support. There is a finite set $A$ of feasible \emph{actions} the receiver can choose from. If the receiver chooses $a \in A$, the payoff to the sender and receiver are given by $u_{s}(a, \omega)$ and $u_{r}(a, \omega)$, respectively, when the state is $\omega$. 
	
	Let $M$ denote a finite set of \emph{messages} (with $|M| \geq \min\{|\Omega|, |A|\})$. In a Bayesian persuasion problem, the sender chooses and commits to a \emph{statistical experiment} $\pi$, which is a mapping from $\Omega$ to $\Delta(M)$. Let $\pi(m|\omega)$ denote the probability of sending message $m$ in state $\omega$ under the experiment $\pi$. Notice a statistical experiment $\pi$ induces a \emph{joint prior} $p_{\pi} \in \Delta(\Omega \times M)$ such that 
	\begin{equation*}
	p_{\pi}(\omega\times m)= \pi(m|\omega)p(\omega), \quad \forall \omega \in \Omega, m \in M.
	\end{equation*}

	Under the ex-ante formulation, the receiver observes the sender's commitment to a statistical experiment and chooses a  \emph{message-contingent action plan}. Let $f: M \rightarrow \Delta(A)$ denote a generic contingent plan and let $f(m)(a)$ denote the probability of taking action $a \in A$ after message $m \in M$ realizes. Let $\mathcal{F} $ denote the set of all contingent plans endowed with the product topology. Given a statistical experiment $\pi$, the receiver's ex-ante expected payoff from choosing the contingent plan $f$ is given by
	\begin{align*}
		U_{r}(\pi, f) = \sum\limits_{m, a ,\omega} f(m)(a)u_{r}(a, \omega)p_{\pi}(\omega \times m).
	\end{align*}
	Similarly, the sender's payoff is given by
	\begin{align*}
		U_{s}(\pi, f) = \sum\limits_{m, a ,\omega} f(m)(a)u_{s}(a, \omega)p_{\pi}(\omega \times m).
	\end{align*}
	Formally, the timing of the (Bayesian) persuasion game is given as follows: 
	\begin{enumerate}
		\item Sender chooses and commits to a statistical experiment. 
		\item Receiver observes the sender's commitment and chooses a contingent plan. 
		\item Nature draws the state, message, and action according to the prior and players' strategies and payoff realizes. 
	\end{enumerate}
	
	\bigskip 	
	The sender's Bayesian persuasion program is defined by\footnote{Following the standard assumption as \cite{kamenica2011bayesian}, this formalization assumes the sender's preferred equilibrium.} 
	\begin{align*}
        \begin{aligned}
		& \max\limits_{\pi, f} &&U_{s}(\pi, f),\\
		&\text{subject to }  &&f \in \arg\max_{f \in \mathcal{F}} U_{r}(\pi, f).
        \end{aligned}
	\end{align*}
	
	In an ambiguous persuasion game, in addition to statistical experiments, the sender can also choose and commit to an \emph{ambiguous experiment}. As defined in BLL, an ambiguous experiment $\Pi$ is a closed and convex set of statistical experiments. Moreover, neither the sender nor the receiver knows which statistical experiment is going to be used to generate the messages. As a result, after the sender commits to an ambiguous experiment and the receiver chooses a contingent plan, both players need to evaluate their expected payoffs in the presence of ambiguity.\footnote{It is important that the sender has no control over which statistical experiment is used. Otherwise, the receiver may be able to infer which experiment will be used from the sender's preference. To achieve this, the sender can either delegate the choice of the statistical experiment to an unknown and payoff-irrelevant third party or let it depend on an exogenously ambiguous event, say the draw from an Ellsberg urn.} 
	
	For now, both players are assumed to be ambiguity averse and represented by the Maxmin Expected Utility (MEU) model: As each statistical experiment induces a joint prior $p_{\pi}$, an ambiguous experiment $\Pi$ induces a set of joint priors, 
	\begin{equation*}
		C_{\Pi} := \{ p_{\pi} \in \Delta(\Omega \times M): \pi \in \Pi \},
	\end{equation*}
	which is a closed and convex subset of $\Delta(\Omega \times M)$. Both the sender and receiver are assumed to take the set $C_{\Pi}$ as the set of joint priors they deem relevant for evaluating their payoffs under MEU. Then given an ambiguous experiment $\Pi$, the receiver's ex-ante payoff from choosing the contingent plan $f$ is 
	\begin{align*}
		U^{MEU}_{r}(\Pi, f) = \min\limits_{\pi \in \Pi} U_{r}(\pi, f) = \min\limits_{p_{\pi} \in C_{\Pi}}\sum\limits_{m, a ,\omega} f(m)(a)u_{r}(a, \omega)p_{\pi}(\omega \times m).
	\end{align*}
	
	Similarly, the sender's ex-ante payoff from the ambiguous experiment $\Pi$ when the receiver chooses the contingent plan $f$ is 
	\begin{align*}
		U^{MEU}_{s}(\Pi, f) = \min\limits_{\pi \in \Pi} U_{s}(\pi, f) = \min\limits_{p_{\pi} \in C_{\Pi}}\sum\limits_{m, a ,\omega} f(m)(a)u_{s}(a, \omega)p_{\pi}(\omega \times m).
	\end{align*}
	
	The timing of the ambiguous persuasion game is analogously the same as the Bayesian persuasion game except that the sender can choose and commit to an ambiguous experiment and if this is the case, Nature also needs to draw a statistical experiment ambiguously. 
	
	The MEU sender's ambiguous persuasion program when facing an MEU receiver is defined by 
	\begin{align*}
        \begin{aligned}
		& \max\limits_{\Pi, f} &&U^{MEU}_{s}(\Pi, f),\\
		&\text{subject to }  &&f \in \arg\max\limits_{f \in \mathcal{F}} U_{r}^{MEU}(\Pi, f). 
        \end{aligned}
	\end{align*}
	
	Note that this program is in general different from program (3) in BLL as 
	\begin{equation*}
		f \in \arg\max_{f \in \mathcal{F}} U_{r}^{MEU}(\Pi, f)
	\end{equation*}
	does not imply $f(m)$ remains optimal when the receiver applies full Bayesian updating to update the prior conditioning on the message $m$. 
	
	\section{No Gains from Ambiguous Persuasion}\label{nogain}
	The main result of this paper establishes that, under the ex-ante formulation of the persuasion game, there is no benefit to the sender from strategically introducing ambiguity into their communication. 
	
	\begin{theorem}\label{ex-ante}
		When both players are MEU decision-makers, the sender's maximum payoff from the ambiguous persuasion program coincides with their maximum payoff from the Bayesian persuasion program. 
	\end{theorem}

	Theorem \ref{ex-ante} delivers a very strong message: when both players are MEU decision makers and ambiguity is introduced without inducing dynamically inconsistent choices, then there will be no benefit to the sender. In other words, all the benefits from ambiguous persuasion identified in BLL under the interim formulation stem from exploiting the receiver's dynamic inconsistency under ambiguity aversion. 

	Before presenting the full proof of Theorem \ref{ex-ante}, I first outline its key steps. Given an ambiguous experiment $\Pi$, the receiver solves a maxmin program for which the minimax theorem applies and saddle points exist. As a result, any contingent plan $f^{*}$ that is a best response to $\Pi$ must also be a best response to some $\pi^{*} \in \Pi$. This implies that the sender is weakly better off committing to $\pi^{*}$ rather than to $\Pi$, since doing so induces the same response from the receiver while weakly improving their own payoff as an MEU decision-maker. 

	\subsection{Proof of Theorem \ref{ex-ante}}\label{proof-ex-ante}
	\begin{proof}[Proof of Theorem \ref{ex-ante}]
		Suppose the sender commits to some ambiguous experiment $\Pi$. Then $f$ can be induced by $\Pi$ if and only if
		\begin{equation*}
			f \in \arg\max\limits_{f \in \mathcal{F}} \min\limits_{p_{\pi} \in C_{\Pi}}\sum\limits_{m, a ,\omega} f(m)(a)u_{r}(a, \omega)p_{\pi}(\omega \times m),
		\end{equation*}
		i.e., $f$ is the solution to the maxmin program. Notice the objective function for the maxmin program is linear and thus continuous in $f$ and $p_{\pi}$. Moreover, $C_{\Pi}$ is a closed and convex subset of a compact set $\Delta(\Omega \times M)$, thus also compact. $\mathcal{F}$ is a finite Cartesian product of the convex and compact set $\Delta(A)$, thus is also convex and compact under the product topology. Therefore, all the conditions for Sion's minimax theorem \citep*{sion1958general} hold and implies that
		\begin{equation*}
			\max\limits_{f \in \mathcal{F}} \min\limits_{p_{\pi} \in C_{\Pi}}\sum\limits_{m, a ,\omega} f(m)(a)u_{r}(a, \omega)p_{\pi}(\omega \times m) =  \min\limits_{p_{\pi} \in C_{\Pi}}\max\limits_{f \in \mathcal{F}}\sum\limits_{m, a ,\omega} f(m)(a)u_{r}(a, \omega)p_{\pi}(\omega \times m),
		\end{equation*}
		i.e., a saddle value for the program exists. Compactness of $\mathcal{F}$ and $C_{\Pi}$ further guarantees the existence of saddle points (see Theorem 8.2 in \citet{aubin2002optima}, for example). Therefore, any solution $f^{*} $ must be part of a saddle point, i.e., there exists some $\pi^{*} \in \Pi$ such that $f^{*}$ is also a best response to $\pi^{*}$. 
		
		When the sender commits to this ambiguous experiment $\Pi$ which induces $f^{*}$, one has
		\begin{align*}
			U^{MEU}_{s}(\Pi, f) & = \min\limits_{p_{\pi} \in C_{\Pi}}\sum\limits_{m, a ,\omega} f^{*}(m)(a)u_{s}(a, \omega)p_{\pi}(\omega \times m) \\
			& \leq \sum\limits_{m, a ,\omega} f^{*}(m)(a)u_{s}(a, \omega)p_{\pi^{*}}(\omega \times m) = U_{s}(\pi^{*}, f^{*}), 
		\end{align*}
		i.e., is weakly worse than committing to the statistical experiment $\pi^{*}$ that also induces $f^{*}$. Therefore, the sender's payoff from any ambiguous experiment is always weakly dominated by a statistical experiment, which implies the optimum of the two programs must coincide. 
	\end{proof}

	\subsection{On the Robustness of Theorem \ref{ex-ante}}\label{robustness}
	Theorem \ref{ex-ante} is established under the assumption that (i) both players share a common unambiguous belief $p$ over the states, and (ii) both are MEU decision makers. This section examines the extent to which the conclusion of Theorem \ref{ex-ante} continues to hold when these assumptions are relaxed.

	The proof of Theorem \ref{ex-ante} reveals that the no-gain conclusion hinges crucially on two key forces. First, any contingent plan that the receiver deems optimal for some ambiguous experiment must also be optimal for a possible statistical experiment. Second, under MEU, the sender evaluates the receiver's contingent plan based on the worst possible statistical experiment, making the sender weakly worse off compared to simply using the statistical experiment (in the set) that induces the same plan. The next theorem shows how these two forces remain valid when relaxing the assumptions of Theorem \ref{ex-ante}. 

	\begin{theorem}\label{thm:robustness}
		The conclusion of Theorem \ref{ex-ante} continues to hold when
		\begin{enumerate}[(i)]
			\item the players have heterogeneous beliefs over the states,
		\end{enumerate}
		and/or
		\begin{enumerate}[(i)] \setcounter{enumi}{1}
			\item the receiver is an Uncertainty Averse decision maker as defined in \cite{cerreia2011uncertainty}.
		\end{enumerate}
		However, the conclusion does not hold when the sender is not an MEU decision maker. 
	\end{theorem}

	A formalization of the statements in Theorem \ref{thm:robustness} requires additional definitions and notations thus are relegated to Appendix \ref{robustness-proof} together with the proof. 

	To build some intuition of Theorem \ref{thm:robustness}, first observe that the two key forces underlying the proof of Theorem \ref{ex-ante} are independent of the players' prior beliefs, as long as they consider the same ambiguous experiment, which is a standing assumption throughout the paper. Moreover, the first force continues to hold as long as the receiver's decision problem takes the form of a maxmin program. This is true when the receiver is an Uncertainty Averse decision maker as defined in \cite{cerreia2011uncertainty}, which includes most models of ambiguity averse preferences in the literature, e.g., MEU, ambiguity-averse smooth ambiguity preferences \citep{klibanoff2005smooth} and variational preferences \citep{maccheroni2006}. Under these relaxed assumptions, the same no-gain conclusion can be established when the sender remains an MEU decision maker.
	
	However, the MEU assumption on the part of the sender is essential for the second force to work. Once the sender departs from MEU, the same argument no longer necessarily holds. In particular, when the aforementioned $\pi^{*}$ is the worst-case experiment for the sender, a non-MEU sender may evaluate $\Pi$ more optimistically and obtain a strictly higher payoff than under the MEU evaluation. In that case, committing to the ambiguous experiment $\Pi$ can strictly outperform committing to $\pi^{*}$. To see this more concretely, consider the following example:

	\begin{example}\label{exp}
		Suppose there are two equally likely states of the world: $\{\omega_{1}, \omega_{2}\}$. The receiver has three feasible actions: $\{a,b,c\}$. The sender's payoff depends only on the action: $u_{s}(c) = 2$, $u_{s}(b) = 1$, and  $u_{s}(a) = 0$. The receiver's payoff is given by the following payoff matrix:  
		\begin{center}
			\begin{tabular}{ccc}
				& $\omega_{1}$ & $\omega_{2}$  \\
				\hline
				a& 2 & -2 \\
				b & 3/2& -1/2  \\
				c & 0 & 0\\
			\end{tabular}
		\end{center}
		There are two messages $\{m_{1}, m_{2}\}$, and let a contingent plan $f$ be denoted by $f(m_{1})f(m_{2})$. 
	\end{example}
	
	Let $\pi^{*}$ denote the optimal statistical experiment which generates posteriors (in terms of the probability of $\omega_{2}$) $1/4$ and $3/4$ with equal probabilities. Given this experiment, the receiver takes action $b$ and $c$ at the two posteriors respectively. The sender's ex-ante payoff is
	\begin{equation*}
		U_{s}(\pi^{*}, bc) = \frac{1}{2} u_{s}(b) + \frac{1}{2} u_{s}(c) = \frac{3}{2}. 
	\end{equation*}
	Consider a statistical experiment $\pi_{\epsilon}$ which generates posteriors $(1/4 - \epsilon)$ and $3/4$ for some $\epsilon > 0$. Bayes plausibility implies that the first posterior is generated with a probability of $1/(2+4\epsilon)$, strictly less than $1/2$. Hence, if fixing the receiver's contingent plan, the sender's ex-ante payoff will be strictly higher under $\pi_{\epsilon}$ compared with $\pi^{*}$: 
	\begin{equation*}
		U_{s}(\pi_{\epsilon}, bc) = \frac{1}{2+4 \epsilon} u_{s}(b) + \frac{1+4\epsilon}{2+4 \epsilon} u_{s}(c) = \frac{3 + 8\epsilon}{2+4\epsilon} > \frac{3}{2} = U_{s}(\pi^{*}, bc).
	\end{equation*}

	Let the ambiguous experiment $\Pi$ be the closed and convex hull of $\{\pi_{\epsilon}, \pi^{*} \} $ and notice that 
	\begin{equation*}
		bc \in \arg\max\limits_{f \in \mathcal{F}} U^{MEU}_{r}(\Pi, f),
	\end{equation*}
	since $\pi^{*}$ minimizes $U_{r}(\pi, bc)$ over all experiments in $\Pi$. Therefore, $bc$ can be induced when the sender commits to $\Pi$.\footnote{For this $\Pi$, $ac$ is also a best response for the receiver. To address potential tie-breaking concerns, the sender can slightly perturb $\pi^{*}$ to generate posteriors $1/4 - \eta$ and $3/4$ for an arbitrarily small $\eta > 0$. This adjustment ensures that $bc$ becomes the receiver's unique best response, while the sender's payoff remains arbitrarily close to that under $\Pi$.} 

	On the other hand, $\pi^{*}$ turns out to be also the minimizer of $U_{s}(\pi, bc)$ among all experiments in $\Pi$, that is,
	\begin{equation*}
		U_{s}^{MEU}(\Pi, bc) = U_{s}(\pi^{*}, bc).
	\end{equation*}
	This confirms that the sender cannot benefit from ambiguous persuasion when they are an MEU decision-maker. However, if the sender is not MEU, it may be the case that $U_{s}(\Pi, bc) > U_{s}^{MEU}(\Pi, bc)$, yielding a strictly higher payoff than the optimal Bayesian persuasion. This occurs, for instance, when the sender has smooth ambiguity preferences or $\alpha$-MEU preferences. For senders with such preferences facing an MEU receiver, a characterization of the optimal payoff from ambiguous persuasion can be found in \cite{cheng2023}.
	
	\section{Relevance of the Ex-Ante Formulation}\label{discussion}
	
	This paper studies ambiguous persuasion under an ex-ante formulation and shows that the results differ significantly from those in BLL's interim formulation. In Bayesian persuasion, the two formulations are equivalent because the receiver maximizes expected utility and updates beliefs using Bayes' rule, and is therefore always dynamically consistent. More fundamentally, this consistency arises from the fact that the receiver's preferences under Bayesian persuasion satisfy Savage's sure-thing principle \citep{savage1972foundations, ghirardato2002revisiting}. In contrast, ambiguity-averse preferences necessarily violate the sure-thing principle, and as a result, the receiver's ex-ante and interim optimal actions under ambiguous persuasion may differ. Therefore, when it is a priori unclear which optimal action the receiver will take, it is essential to consider both perspectives. On top of that, there are also settings in which the ex-ante formulation may be arguably more relevant. The following presents three such settings.

	\paragraph{Timing of the Game Follows the Ex-ante Formulation.} In the ex-ante formulation, the receiver chooses a message-contingent action plan before observing the message and cannot change it thereafter. This setup naturally arises when, for example, the sender is constrained to communicate through a contingent contract that specifies states and actions, and the receiver is legally required to sign at the ex-ante stage. This structure is common in many real-world applications, including insurance policies, earn-out agreements in mergers and acquisitions, as well as contingency clauses in home purchase agreements. In such settings, the receiver would agree to the contract only if the prescribed action plan is ex-ante optimal, making the ex-ante formulation the more relevant approach. 

	\paragraph{The Receiver Chooses Updating Rules That Ensure Dynamic Consistency.} Because ambiguity averse preferences violate the sure-thing principle, selecting an updating rule becomes an active and non-trivial decision for the receiver. Among the many possible updating rules, \cite{hanany2007updating, hanany2009updating} characterize a family of updating rules that guarantee dynamic consistency, that is, they ensure the receiver's ex-ante optimal action remains interim optimal after updating. If the receiver adopts such an updating rule, then there is no distinction between the interim and ex-ante formulations, and both are consistent with the results presented in this paper. The experiment by \cite{SHISHKIN2023105610} mentioned in the introduction presents evidence that many subjects indeed choose such updating rules when facing an ambiguous experiment. 
	
	\paragraph{The Receiver is Sophisticated and Able to Tie Their Own Hand.} Even if the receiver does not adopt an updating rule that guarantees dynamic consistency, they may still be sophisticated enough to anticipate that they will behave inconsistently under ambiguity. As a result, such a receiver at the ex-ante stage of the game may strictly prefer to ``tie their own hand'' by committing to the ex-ante optimal action plan.\footnote{A behavioral characterization of preferences that value such commitments in decision-making is given by \cite{siniscalchi2011dynamic}.} Recall the ex-ante stage here refers to the point after observing the sender's commitment to an experiment but before learning the realized message. In other words, this form of ``commitment'' by the receiver only regulates their own behavior and is not a strategic move that could affect the sender's choice of experiment. To tie their own hand and implement the ex-ante optimal action, the receiver may, for instance, pre-program an algorithm to determine their actions or delegate the decision to a third party. In these cases, the ex-ante formulation is again the more appropriate framework for analyzing the game.

	\bibliographystyle{econ}
	\bibliography{references}

	\numberwithin{equation}{section}
	\numberwithin{theorem}{section}
	\numberwithin{lemma}{section}

	\begin{appendices}	

	\section{The Formalization and Proof of Theorem \ref{thm:robustness}}\label{robustness-proof}
	This section formalizes the statements in Theorem \ref{thm:robustness} by introducing two alternative setups of the persuasion game, each corresponds to the relaxation of one of the original assumptions. The first setup allows the players to hold heterogeneous beliefs over the states. The second setup permits the receiver to have more general Uncertainty Averse preferences. The case combining both relaxations is straightforward and thus is omitted.

	\subsection{Heterogeneous Beliefs over States}
	Let $p_{s}$ and $p_{r}$ denote the sender and receiver's priors over the states, respectively. The sender's program when choosing among only statistical experiments is given by
	\begin{align}\label{statistical}
		\begin{aligned}
		& \max\limits_{\pi, f} && \sum_{\omega,m,a}f(m)(a)u_{s}(a, \omega)p_{s}(\omega)\pi(m | \omega),\\
		&\text{subject to }  &&f \in \arg\max_{f \in \mathcal{F}} \sum_{\omega,m,a}f(m)(a)u_{r}(a, \omega)p_{r}(\omega)\pi(m | \omega).
		\end{aligned}
	\end{align}

	Similarly, when the sender is allowed to choose ambiguous experiments, the sender's program becomes
	\begin{align}\label{ambiguous}
		\begin{aligned}
		& \max\limits_{\Pi, f} && \min\limits_{\pi \in \Pi} \sum_{\omega,m,a}f(m)(a)u_{s}(a, \omega)p_{s}(\omega)\pi(m | \omega),\\
		&\text{subject to }  &&f \in \arg\max\limits_{f \in \mathcal{F}}  \min\limits_{\pi \in \Pi}\sum_{\omega,m,a}f(m)(a)u_{r}(a, \omega)p_{r}(\omega)\pi(m | \omega).
		\end{aligned}
	\end{align}
	
	Then the first statement of Theorem \ref{thm:robustness} is formalized by the following lemma.

	\begin{lemma}\label{lem:ambiguity-states}
		The sender's optimal payoff from program \eqref{statistical} coincides with their optimal payoff from program \eqref{ambiguous}.
	\end{lemma}
	
	\begin{proof}[Proof of \ref{lem:ambiguity-states}]
		Suppose the sender commits to some ambiguous experiment $\Pi$. Then $f$ can be induced by $\Pi$ if and only if
		\begin{equation*}
			f \in \arg\max\limits_{f \in \mathcal{F}} \min\limits_{\pi \in \Pi}\sum_{\omega,m,a}f(m)(a)u_{r}(a, \omega)p_{r}(\omega)\pi(m | \omega).
		\end{equation*}

		By the same argument, the minimax theorem applies and saddle points exist. Therefore, an optimal $f^{*}$ must be a saddle point such that it is optimal under some $\pi^{*} \in \Pi$. Next,the sender's payoff from committing to $\Pi$ satisfies
		\begin{align*}
			U_{S}(\Pi, f^{*}) & =  \min\limits_{\pi \in \Pi} \sum_{\omega,m,a}f^{*}(m)(a)u_{s}(a, \omega)p_{s}(\omega)\pi(m | \omega) \\
							& \leq \sum_{\omega,m,a}f^{*}(m)(a)u_{s}(a, \omega)p_{s}(\omega)\pi^{*}(m | \omega) = U_{S}(\pi^{*}, f^{*}).
		\end{align*}
		Therefore, the sender's payoff from any ambiguous experiment is weakly dominated by a statistical experiment, which implies the optimum of the two programs must coincide.
	\end{proof}

	\subsection{Uncertainty Averse Receiver}
	The most general family of preferences that displays ambiguity aversion is the Uncertainty Averse preferences, defined and axiomatized by \citet{cerreia2011uncertainty}. According to the Uncertainty Averse representation, the receiver's ex-ante payoff from choosing a plan $f$ when the sender commits to an ambiguous experiment $\Pi$ is given by 
	\begin{equation*}
		U^{UAP}_{r}(\Pi, f) = \min\limits_{p \in \Delta(\Omega \times M)} G_{\Pi}\left( \sum\limits_{m, a, \omega} f(m)(a)u_{r}(a,\omega)p(\omega \times m), p \right),
	\end{equation*}
	for some function $G_{\Pi}: T \times \Delta(\Omega \times M) \rightarrow (-\infty, +\infty]$ with $T = [\min_{a,\omega} u_{r}(a, \omega), \max_{a,\omega} u_{r}(a,\omega)]$ and satisfies: 
	\begin{enumerate}[(i)]
		\item $G_{\Pi}(\cdot, \cdot)$ is lower semi-continuous and quasi-convex. 
		\item $G_{\Pi}(\cdot, p)$ is increasing for all $p \in \Delta(\Omega \times M)$. 
		\item $\min_{p \in \Delta(\Omega \times M)} G_{\Pi}(t, p) = t$ for all $t \in T$. 
		\item $G_{\Pi}(\cdot, p)$ is extended-valued continuous on $T$ for all $p \in \Delta(\Omega \times M)$.  
		\item $C_{\Pi} = \{p \in \Delta(\Omega \times M) : G_{\Pi}(\min_{a,\omega} u_{r}(a, \omega), p) < \infty\}$, which is closed and convex as $G_{\Pi}(\cdot, \cdot)$ is lower semi-continuous and quasi-convex. 
	\end{enumerate}
	
	Notice (i) - (iv) are standard conditions for a representation of Uncertainty Averse preferences.\footnote{See \citet{cerreia2011uncertainty} for the required axioms and discussions.} (v) is the additional requirement that the receiver views $C_{\Pi}$ as the set of relevant joint priors according to their Uncertainty Averse preferences. In other words, $G_{\Pi}$ is possibly finite only under those joint priors in $C_{\Pi}$. Notice that if the receiver has MEU preferences (as a special case of Uncertainty Averse preferences), this requirement simply reduces to the previous requirement on the receiver's set of joint priors.\footnote{Also see Section 5.2.2 in \citet{cerreia2011uncertainty} for the discussion in the case of smooth ambiguity preferences \citep{klibanoff2005smooth}.} 

	On the other hand, the sender remains an MEU decision maker. The MEU sender's ambiguous persuasion program when facing an Uncertainty Averse receiver is defined by
	\begin{align*}
        \begin{aligned}
		& \max\limits_{\Pi, f} &&U^{MEU}_{s}(\Pi, f),\\
		&\text{subject to }  &&f \in \arg\max\limits_{f\in \mathcal{F}} U^{UAP}_{r}(\Pi, f).            
    \end{aligned}
	\end{align*}
	
	Then the second statement of Theorem \ref{thm:robustness} is formalized by the following lemma.

	\begin{lemma}\label{thm_ua}
		When the sender is an MEU decision maker and the receiver is an Uncertainty Averse decision maker, the sender's maximum payoff from the ambiguous persuasion program coincides with the maximum payoff from the Bayesian persuasion program. 
	\end{lemma}
			
	\begin{proof}[Proof of Lemma \ref{thm_ua}]
		The proof here is similar to the proof of Lemma 2 in \cite{LI2020105012}. Suppose the sender commits to some ambiguous experiment $\Pi$. Then $f$ can be induced by $\Pi$ if and only if
		\begin{equation*}
			f \in \arg \max\limits_{f \in \mathcal{F}}\min\limits_{p \in C_{\Pi}} \tilde{G}_{\Pi}(f, p) : = \arg\max\limits_{f \in \mathcal{F}}\min\limits_{p \in C_{\Pi}}G_{\Pi}\left( \sum\limits_{m, a, \omega} f(m)(a)u_{r}(a,\omega)p(\omega \times m), p \right).
		\end{equation*}
		Both $\mathcal{F}$ and $C_{\Pi}$, as argued in the proof of Theorem \ref{ex-ante}, are compact convex subsets of linear topological spaces. For each $p \in C_{\Pi}$, $\tilde{G}_{\Pi}(\cdot, p)$ is continuous following from continuity of $G_{\Pi}(\cdot, p)$ and linearity of $\sum_{m, a, \omega} f(m)(a)u_{r}(a,\omega)p(\omega \times m)$. $\tilde{G}_{\Pi}(\cdot, p)$ is quasi-concave following from monotonicity of $G_{\Pi}(\cdot, p)$ and linearity of $\sum_{m, a, \omega} f(m)(a)u_{r}(a,\omega)p(\omega \times m)$. For each $f \in \mathcal{F}$, $\tilde{G}_{\Pi}(f, \cdot)$ is lower semi-continuous and quasi-convex following from condition (i). Thus all the conditions for Sion's minimax theorem hold and thus imply that
		\begin{align*}
			&\max\limits_{f \in \mathcal{F}}\min\limits_{p \in C_{\Pi}}\tilde{G}_{\Pi}(f, p) =  \min\limits_{p \in C_{\Pi}}\max\limits_{f \in \mathcal{F}} \tilde{G}_{\Pi}(f, p).
		\end{align*} 

		Again, it further implies that any solution $f^{*}$ must be part of a saddle point, i.e., there exists some $p^{*} \in C_{\Pi}$ such that
		\begin{equation*}
			f^{*} \in \max\limits_{f \in \mathcal{F}}\tilde{G}_{\Pi}(f, p^{*})  = \arg\max\limits_{f \in \mathcal{F}}G_{\Pi}\left( \sum\limits_{m, a, \omega} f(m)(a)u_{r}(a,\omega)p^{*}(\omega \times m), p^{*} \right).
		\end{equation*}	
		Then monotonicity of $G_{\Pi}(\cdot, p)$ implies that 
		\begin{align*}
			& \tilde{G}_{\Pi}(f^{*}, p^{*}) \geq \tilde{G}_{\Pi}(f, p^{*})\\
			\Leftrightarrow & \sum\limits_{m, a, \omega} f^{*}(m)(a)u_{r}(a,\omega)p^{*}(\omega \times m) \geq \sum\limits_{m, a, \omega} f(m)(a)u_{r}(a,\omega)p^{*}(\omega \times m), 
		\end{align*}
		i.e., $f^{*}$ is also optimal for the receiver against $p^{*}$ with $p_{\pi^{*}} = p^{*}$. From this point on, exactly the same argument as in the proof of Theorem \ref{ex-ante} applies. 
	\end{proof}	

\end{appendices}
\end{document}